\documentclass[12pt,preprint]{aastex}         
\usepackage{graphicx}
\usepackage{epstopdf}
\def\be{\begin{equation}}
\def\ee{\end{equation}}

\def\tt{\mbox{\boldmath $\theta $}}

\def\lsim{\lower 2pt \hbox{$\, \buildrel {\scriptstyle <}\over
         {\scriptstyle \sim}\,$}}

\begin{document}
\newcommand{\figureout}[3]{\psfig{figure=#1,width=5.5in,angle=#2} 
   \figcaption{#3} }

\title{PULSAR PAIR CASCADES IN A DISTORTED MAGNETIC DIPOLE FIELD}

\author{Alice K. Harding\altaffilmark{1} \& Alex G. Muslimov\altaffilmark{1,2}} 
  
\altaffiltext{1}{Astrophysics Science Division,      
NASA/Goddard Space Flight Center, Greenbelt, MD 20771}

\altaffiltext{2}{Universities Space Research Association/CRESST, Columbia, MD 21044}


\begin{abstract}
We investigate the effect of a distorted neutron star dipole magnetic field on pulsar pair cascade multiplicity and pair death lines.  Using a simple model for a distorted dipole field that produces an offset polar cap, we derive the accelerating electric field above the polar cap in space charge limited flow.  
We find that even a modest azimuthally asymmetric distortion can significantly increase the accelerating electric field on one side of the polar cap and, combined with a smaller field line radius of curvature, leads to larger pair multiplicity.  The death line for
producing pairs by curvature radiation moves downward in the $P$-$\dot P$ diagram, allowing high pair multiplicities in a larger percentage of the radio pulsar population.  
These results could have important implications for the radio pulsar population, high energy pulsed emission and the pulsar contribution to cosmic ray positrons.
\end{abstract}

\keywords{pulsars: general --- stars: neutron}

\pagebreak
  
\section{INTRODUCTION}

Rotation-powered pulsars are thought to produce electron-positron pairs through electromagnetic cascades in their magnetospheres (Sturrock 1971).  
Such pair cascades may occur both in the magnetic polar regions, where high energy photons radiated by accelerated particles  undergo pair conversion in the strong near-surface magnetic fields (Ruderman \& Sutherland 1975, Daugherty \& Harding 1982) or in outer gaps, where high energy photons interact with thermal X-rays from the neutron star (NS) surface 
(Cheng, Ho \& Ruderman 1986, Hirotani 2008).  For some time though, there has been observational evidence that the number of pairs that can be generated in pair cascades 
by the existing standard models of particle acceleration in magnetic dipole fields is insufficient to account for the optical to X-ray emission from the synchrotron nebulae powered by the pulsars.  Estimates of the 
pair multiplicity (the number of pairs produced by each primary accelerated particle) of about $10^5 - 10^6$ needed to account for the emission from the Crab pulsar wind nebula (PWN) 
(DeJager et al. 1996) of about $10^5$ for the Vela PWN (DeJager  2007) are more than one order of magnitude larger than the theoretical pair multiplicities.  Synchrotron 
absorption models for the eclipse in the double pulsar system PSR J0737-3039 (Arons et al. 2005, Lyutikov 2004) require a pair multiplicity of around $10^6$ for the recycled 
22 ms pulsar in that system.
There has also been a long-standing problem in understanding how long-period pulsars are able to produce coherent radio emission, thought to require electron-positron pairs
( Arons 1983).  Again, the standard pulsar models are not able to account for the operation of robust pair cascades in these aging pulsars (Hibschmann \& Arons 2001).
More recently, the discovery of pulsed gamma-ray emission from a large number of millisecond pulsars by the  Large Area Telescope (LAT) on the 
$Fermi$ Gamma-Ray Space Telescope (Abdo et al. 2009) has revealed light curves that are best modeled by 
narrow radiation gaps in the outer magnetosphere (Venter et al. 2009, Abdo et al. 2010).  Such narrow gaps require screening of the accelerating electric field over most of
the magnetosphere by a pair multiplicity that is orders of magnitude larger than standard models are able to produce (Harding \& Muslimov 2002).

This motivates us to investigate the effect of a large-scale non-dipolar NS magnetic field geometry on pulsar pair cascades. 
Off-centered dipole-like magnetic fields seem to be prevalent among planets in the Solar System (Russell \& Dougherty 2010), and in some stars.  For example, the magnetic fields of isolated magnetic white dwarfs may possess an off-centered dipole or dipole plus quadrupole configurations (see e.g. Putney \& Jordan 1995) which is consistent with the idea that the magnetic $Ap$ and $Bp$ stars, the predecessors of highly magnetic white dwarfs, favor an off-centered dipole field.  However, it is very likely that the surface magnetic fields of white dwarfs are much more complex (Wickramasinghe 2001).  Recent modeling of X-ray pulse profiles from millisecond pulsars (MSPs) shows evidence for offset dipole fields or offset polar caps in PSR J0437-4715 (Bogdanov et al. 2007) and PSR J0030+0451 (Bogdanov \& Grindlay 2009).  
Modeling of the X-ray pulsations of the NSs in Low-Mass X-ray Binaries, the progenitors of millisecond pulsars, show possible evidence of even more extreme 
magnetic field distortions (Lamb et al. 2009), that could result from distortion of the global magnetic field by e.g. the crustal plate tectonics (Ruderman, 1991). 
In pulsar theory, non-dipolar magnetic fields have been discussed by many authors and long before the $Fermi$ era.   Arons (1983) argued that pair creation in long period ($P > 1$ s) pulsars was difficult and that distortions of polar field lines due to large-scale, non-dipolar surface fields of magnitude comparable to the dipole component, and of much smaller radius of curvature, could enable pair creation in older pulsars.  Arons (1997) concluded that ``while the frame dragging effect in star centered dipole geometry does improve comparison of the theory with observation, an unacceptably large fraction of the observed stars outside the bounds of pair creation theory still persists".  He conjectured that dipole offsetting improves the correspondence between theory and observation, and could account for the ``death valley" observed for pulsars 
approaching the radio death line.  In this Letter we report our preliminary assessment of the effect of a distorted dipole magnetic field of a NS on the efficiency of pair creation. A more detailed treatment will appear in a later publication.

\section{Magnetic and Electric Field of a Distorted Dipole}

We propose the following heuristic model of an asymmetric magnetic field of a pulsar that can be used to approximate plausible distortions of the magnetic field of a NS. 
We introduce an azimuthal asymmetry to the field lines of an originally symmetric dipole such that the field lines over half of the polar cap (PC) have relatively smaller radius of curvature and over the other half of the PC have larger radius of curvature (see Figure 1 below).  Consequently, one side of the PC is larger and the PC is effectively shifted from the [star] center of symmetry.   
  
In magnetic spherical polar coordinates ($\theta $, $\phi $) the magnetic field is assumed to have the following form:
\be
{\bf B} \approx {B_0\over {\eta ^3}}~\left[ \hat{\bf r}~\cos [\theta (1+a)] +{1\over 2} ~\hat{\tt } \sin [\theta (1+a)] \right],
\label{B}
\ee
where $\eta = r/R_{\rm ns} $ is the dimensionless radial coordinate in units of stellar radius, 
$a$ is the parameter characterizing the distortion of polar field lines. 
Note that vector $\bf B$ is only approximately solenoidal, since we omitted a small $B_{\phi }$-component ($\sim B_{\theta }\partial a/\partial \phi $) and higher-order
corrections in $B_r$ and $B_{\theta }$ that enter only higher-order terms neglected in the formulae below. In subsequent studies we will discuss these terms as well as more general
 types of field distortion.
If the magnetic axis lies in the x-z plane, $a = \epsilon \cos \phi$ produces an effective offset of the PC in the x-z plane and $a = \epsilon \sin \phi $ an effective offset of the PC in the y-z plane, where $0 \leq \epsilon  < 1$.  By using formula (\ref{B}) we get the equation of the magnetic field line, which we give here in the small-angle approximation valid
near the NS surface,
\be
\theta = {\xi \over {1+a}}~x^{(1+a)/ 2}, 
\label{FL}
\ee
where $x = r/r_{\rm LC}$ is the radial distance in units of the light cylinder radius, $r_{\rm LC} = c/\Omega$; and $0\leq \xi \leq 1$ is the colatitude of a footpoint of polar field line normalized by the colatitude of the PC boundary.   Figure 1 shows the field lines computed from Eqn (2) plotted in the x-z plane. 
The field line radius of curvature (in units of $r_{\rm LC}$) is
\be
x_c = {4\over 3\xi}\,x^{(1-\epsilon\cos\phi)/2}\,{1+\alpha\over 1+\beta}
\ee
where
\be 
\alpha = {3\over 8} \Delta, ~~~~~ \beta = {1\over 6} \Delta,  ~~~~~\Delta = \xi^2 x^{1+a}.
\ee

The effective shift of the PC on the NS surface is approximately,
\be
\Delta r_{\rm PC} \simeq R_{\rm ns}\theta_0 \left[1 - \theta_0^{\epsilon}\right],
\ee
where $\theta_0 = (\Omega R_{\rm ns}/c)^{1/2}$ is the canonical half-angle of the PC, $\Omega$ is the pulsar rotation rate and $R_{\rm ns}$ is the NS radius.  Since the effective offset of the PC is a fraction of 
the PC opening angle, it is a very small fraction of the stellar radius  for normal pulsars and somewhat larger fraction for millisecond pulsars. In the rest of the paper we will be referring to $\epsilon $ as the offset parameter.

Neglecting the static general-relativistic corrections, we can write the accelerating electric field, assuming the boundary conditions of space-charge limited flow (SCLF)
(cf. Harding \& Muslimov 1998), as
\be
E_{||} \approx - {1\over 2}\left( {{\Omega R_{\rm ns}}\over c} \right) ^2 B_0 {{x^a}\over {(1+a)^2}} 
\left\{ {\kappa \over {\eta ^4}}[3+ a(\eta ^3 -1)] \cos \chi + {3\over 8}{ {x^{a/2}}\over {(1+a)}}{{\theta _0}\over {\sqrt{\eta }}}\xi \sin \chi \cos \phi \right\} (1-\xi ^2), \label{Epar}
\ee
where $\chi $ is the pulsar obliquity, $\kappa \approx 0.15~I_{45}/R_6^3$ ($I_{45} = I/10^{45}$~g$\cdot $cm$^2$, $R_6 = R_{\rm ns}/10^6$ cm, $I$ is NS moment of inertia) is the parameter accounting for the general-relativistic frame dragging ( Muslimov \& Tsygan 1992).
Note that the second term in Eqn (\ref{Epar}) is significantly simplified and is not valid for very large values of $\chi$.
We also note that the current of primary charges assumed in the SCLF models, which is $j = c\rho_{\rm GJ}(0)$, is not consistent with the polar cap current that has been derived in force-free magnetosphere models (e.g. Contopoulos et al. 1999, Timokhin 2006, Spitkovsky 2006), where $\rho_{\rm GJ}(0) = \Omega B_0/2\pi c$ is the Goldreich-Julian charge density at the surface.  The discharge properties under these conditions were discussed by Beloborodov (2008) and may involve non-steady charge outflow and electric fields.  We use here the SCLF electric field for comparison with the many previous studies of pulsar pair cascades and pair death lines.

\section{Pair Cascade Simulation and Results}

Using the distorted field structure and accelerating electric field described above, we have simulated a full pair cascade above the pulsar polar cap to compare the 
pair multiplicity and pair death lines for different offset parameters $\epsilon$.  The full pair cascade simulation code, described in Daugherty \& Harding (1982) and
Daugherty \& Harding (1996), has been combined with the acceleration, early cascade development and pair front formation used in Harding \& Muslimov (1998).  
The combined code is thus able to follow a primary particle through its acceleration in the parallel electric field, taking into account inverse Compton and curvature radiation
losses, the emission of photons by curvature radiation (CR), the establishment of the pair front (above which the electric field is screened) and the full cascade of pairs and
their synchrotron radiation.  Since the electric field is screened over a very short length scale by CR pairs (Harding \& Muslimov 2001), for the present calculation we
simply assume that the electric field is zero above the pair formation front (PFF).  
Since the results are fairly sensitive to NS equation of state (EoS), we choose for normal pulsars
a mass of $M_{\sun} = 1.45$ and radius $R_{\rm ns} = 10$ km, which give a moment of inertia $I = 1.13 \times 10^{45}\,\rm g\,cm^2$ (Lattimer \& Prakash 2007), and for MSPs, 
that may have accreted mass during their spin-up phase, we choose $M_{\sun} = 2.15$,  $R_{\rm ns} = 9.9$ km and $I = 1.56 \times 10^{45}\,\rm g\,cm^2$ from a rotating NS 
model (Friedman et al. 1986).

The resulting pair multiplicities for a range of pulsar periods and surface magnetic field strengths, and for offset parameters $\epsilon = 0, 0.1, 0.2, 0.4$ and $0.6$ are
shown in Table 1 for the normal pulsar population and in Table 2 for millisecond pulsars.  For these calculations, we have taken $a = \epsilon \sin \phi$ for an
effective offset in the y-z plane and magnetic azimuth angle $\phi = 270^\circ$, where the field distortion produces the most favorable conditions for pair creation, i.e. smallest radius curvature,
largest PC angle and largest electric field.  Pair multiplicities are computed for one representative magnetic colatitude of $\xi = 0.5$.  
The increase in the parallel electric field, in this case on the trailing side of the PC, by a factor of about $\theta_{0}^{2a}/(1+a)^2$, results in a larger voltage at the PFF and is the 
most important effect leading to larger pair multiplicities since the maximum curvature radiation energy is proportional to the cube of the particle Lorentz factor. 
For normal pulsars, even modest values of $\epsilon$ that are equivalent to offsets of a few percent
of the stellar radius lead to orders of magnitude increases in the pair multiplicity.   Small offsets also enable long period pulsars, that cannot produce any pairs 
from CR in a pure dipole field, to produce high-multiplicity pair cascades.  Most MSPs cannot produce pairs from CR in a dipole field but with some offset of the PC 
would be able to produce high pair multiplicity.  The plateaus and peaks in the multiplicity, that can be seen in Table 1 for $B_0 > 3 \times 10^{12}$ G, are the result of 
pairs being produced increasingly in the ground Landau state in high fields (Baring \& Harding 2001) which inhibits the cascade synchrotron radiation.  Photon splitting,
which we have not included in these calculations, would further reduce pair multiplicity above $10^{13}$ G.  Although the multiplicity grows with increasing $\epsilon$ and increasing
field strength, it saturates at a value below $10^5$.  This limit,  also noted by Medin \& Lai (2010), results from pairs being produced with higher and higher average energy at 
smaller angles to the magnetic field as the primary electron energy, and thus the maximum  energy of the CR photons, increases.  

The increase in pair multiplicity from a distorted dipole field can move many pulsars above the death line for robust pair creation.   Figure 2 shows the death lines for 
pair creation by CR in the $P$-$\dot P$ diagram for different offset parameters, computed for the two different EoS for normal and MSPs described above.  The lines plotted 
in Figure 2, determined by the method described in Harding \& Muslimov (2002), show the lowest $\dot P$ required for a pulsar at a given period to establish a PFF.  The death line
for a pure dipole cuts through the middle of the normal pulsar population and through the upper edge of the MSP population.  As $\epsilon$ increases, the lines move 
significantly downward through both populations.  For $\epsilon = 0.4$, nearly all pulsars including the 8.5 s PSR J2144-3933 (Young, Manchester \& Johnston 1999) 
are able to produce pair cascades which could enable 
coherent radio emission.  The lines decrease in slope as they move from longer to shorter periods and from small to large $\epsilon$ as the particle acceleration becomes
more limited by curvature radiation reaction.  For the shortest period MSPs, the acceleration is completely radiation-reaction limited (Luo et al. 2000) which causes a slower increase in 
electron Lorentz factor and impedes the development of pair cascades.  The envelope of the observed pulsar population seems to be consistent with the shape predicted from this limit.

\section{Discussion} 

We have found that distortions of a dipole magnetic field that produce small offsets of the PC can produce large increases in the multiplicity of pulsar pair cascades
and allow pair cascades in pulsars not able to initiate cascades in centered dipole fields.  Previous estimates of the effect of offset vacuum dipoles on pair death lines included 
only the decreased field line radius of curvature, concluding that dipole offsets of (0.7 - 0.8)$R_{\rm ns}$ (Arons 1997) or even 0.95 $R_{\rm ns}$ (Medin \& Lai 2010), 
that are large fractions of a stellar radius, are required.  
Our calculation also takes into account the change in $E_\parallel$ and the particle acceleration energy, which is by far the strongest effect.  
Since this effect is sensitive to effective offsets that are fractions of a PC radius, we
show that many older pulsars and MSPs would be able to sustain robust pair cascades with dipole offsets that are small fractions of a stellar radius.  
Although distortions of the dipole field could be intrinsic to the NS through asymmetries in the interior currents, present either from birth or as a result of spin-up or spin-down
evolution, magnetic dipole fields that are distorted by rotation or currents also produce offset PCs.  The sweepback of field lines of a retarded vacuum dipole (Deutsch 1955) 
causes an offset of the PC towards the trailing side, opposite the direction of rotation (Dyks \& Harding 2004), and the force-free magnetosphere, with a similar sweepback of field lines, has 
even greater offset of the PC (Bai \& Spitkovsky 2010).

We will mention here a number of implications of our results.   
For high enough offsets, the entire radio MSP population could generate pair cascades.  For the offset of $\epsilon \sim 0.6$ implied by the 
apparent shift of $\sim 1$ km of the heated PC of PSR J0437-4715 (Bogdanov et al. 2007), with $P = 5.75$ ms and $\dot P = 1.4 \times 10^{-20}$, a pair cascade 
multiplicity of about $10^2$ is possible (cf. Table 2).  
The ability of MSPs with distorted dipoles to generate high-multiplicity pair cascades that can screen 
the $E_{\parallel}$ over most of the open field could possibly account for the prevalence of sharply-peaked light curves of $\gamma$-ray MSPs that would require 
narrow acceleration gaps.
Higher pair multiplicity in both normal and MSPs could also enhance their contribution to local Galactic cosmic-ray positrons.
However, we find that the saturation in pair cascade multiplicity with increasing particle energy limits the extent to which a dipole distortion can produce even higher multiplicity.
Therefore, even large values of $\epsilon $ cannot produce the levels of multiplicity in young pulsars required to produce synchrotron emission observed from PWNe.  
We find for example that the Crab pulsar, with $P = 0.033$ s and $B_0 \sim 3 \times 10^{12}$ G could only produce
a pair multiplicity up to $\sim 3 \times 10^4$ for $\epsilon = 0.4$ corresponding to an offset from Eqn (3) of  6\% of the stellar radius.  
A pair multiplicity of $\sim 10^6$ in PSR J0737-3039A would also not be achievable with any degree of offset. 

Distorted dipole fields with offset PCs will generally introduce asymmetries in the pulsar emission as well as in the pair multiplicity.  Since the particle acceleration and field line curvature 
vary over the PC, the pair multiplicity will be larger over one half of the PC in the direction of the offset.  Pulsars that are below the death line for centered dipole fields, 
including many MSPs, will have pair cascades only on that side of the PC.  The variation in pair multiplicity across the PC could result in asymmetric radio emission 
depending on how the coherent mechanism is related to the multiplicity.  
The higher accelerating field and voltage on the offset side of the PC will produce asymmetric heating of the PC and thus inhomogeneous thermal X-ray emission.
The structure and energetics of the proposed slot gaps (SG) that form between the boundary of the open magnetic field and the upward curving PFF 
(Arons \& Scharlemann 1979), and can accelerate particles to high altitude (Muslimov \& Harding 2004), could be strongly affected by a distorted dipole.   The particle Lorentz factor
$\gamma$ in the SG , which is expected to reach curvature radiation-reaction limit such that $\gamma \propto E_{\parallel}^{1/4}$, will be larger on one side of the PC,
producing CR emission power proportional to $E_\parallel$ that is larger than for a dipole field.  An azimuthal asymmetry of both the radiation power and width of the SG 
would change both the $\gamma$-ray luminosity and light curves. 
A further application of these results is 
in pulsar population studies.  The decrease in the observed radio pulsar population just above the death line as pulsars spin-down and evolve from left to right across the 
$\dot P$-$P$ diagram has been difficult to explain.  But a range of dipole offsets across the population would produce a spread of pair/radio death lines and thus produce a death valley.  It is interesting that the no-offset death line shown in Figure 1 occurs near the densest part of the normal pulsar population.  We will explore many of these
consequences of increased pair multiplicity in distorted dipole fields in future studies.

\acknowledgments  
AKH thanks the Aspen Center for Physics where fruitful discussions, particularly with J. Arons, A. Timokhin, O. DeJager and A. Spitkovsky, provided stimulation for this work.  We also acknowledge support from the NASA Astrophysics Theory and Fundamental Physics Program, the $Fermi$ Guest Investigator Program and the Universities Space Research Association.

\newpage
\begin{table}
\caption{Pair Cascade Multiplicity for Normal Pulsars} \label{tbl-1}
\begin{center}
\begin{tabular}{lcccccc}
\tableline
$B_0 (10^{12}$ G) & {\large $\epsilon$} & && {Period (s)} &&\\
\tableline
&& 0.033 & 0.1  & 0.3 & 1.0 & 3.0 \\
\tableline
0.33 & 0.0 & 1.48E3 & 5.66E1& 0.0E0 & 0.0E0 & 0.0E0 \\
& 0.1&  2.27E3 & 2.47E2 & 0.0E0 & 0.0E0 & 0.0E0 \\
& 0.2& 3.45E3 & 9.08E2 & 1.3E-5 & 0.0E0 & 0.0E0 \\
& 0.4 & 7.90E3 & 3.19E3 & 9.74E2 & 1.93E1 & 0.0E0 \\
\tableline
1.0 & 0.0 &  4.81E3 & 4.79E2 & 3E-12 & 0.0E0 & 0.0E0 \\
& 0.1 & 8.13E3 & 1.52E3 & 3.44E1 & 0.0E0 & 0.0E0 \\
& 0.2 & 1.15E4 & 3.01E3 & 3.41E2 & 0.0E0 & 0.0E0 \\
& 0.4 &  2.39E4&  9.90E3 & 3.29E3 & 5.72E2 & 5.6E-8 \\
\tableline
3.0 & 0.0 & 1.07E4 & 2.08E3 & 4.71E1 & 0.0E0 & 0.0E0 \\
& 0.1& 1.49E4 & 4.76E3 & 4.42E2 & 0.0E0 & 0.0E0 \\
& 0.2& 1.90E4 & 8.21E3 & 1.49E3 & 3.32E1 & 0.0E0 \\
& 0.4 &  2.64E4 & 1.58E4 & 7.86E3 & 2.39E3 & 1.91E2\\
\tableline
10 & 0.0 & 1.26E4 & 7.90E3 & 1.94E3 & 8E-10 & 0.0E0 \\
& 0.1 & 1.39E4 & 9.97E3 & 4.50E3 & 1.88E2 & 0.0E0 \\
& 0.2 & 1.46E4 & 1.14E4 & 7.19E3 & 1.90E3   & 0.0E0 \\
& 0.4 & 1.53E4 & 1.32E4 & 1.11E4 & 7.87E3 & 3.64E3\\
\tableline     
\end{tabular}
\end{center}
For $\chi = 60^0$, $\phi = 270^0$, $\xi = 0.5$ and neutron star parameters $M_\sun = 1.45$, $R_{\rm ns} = 10$ km and $I = 1.13 \times 10^{45}\,\rm g\,cm^2$.
\end{table}
     
\newpage
\begin{table}
\caption{Pair Cascade Multiplicity for Millisecond Pulsars} \label{tbl-1}
\begin{center}
\begin{tabular}{lcccc}
\tableline
$B_0 (10^{9}$ G) & {\large $\epsilon$} & & {Period (ms)} &\\
\tableline
&& 2 & 5  & 10  \\
\tableline
0.3 & 0.0 &  4.4E-6 & 0.0E0 & 0.0E0  \\
& 0.1&3.3E-1 & 0.0E0 & 0.0E0 \\
& 0.2& 2.16E0  & 0.0E0 & 0.0E0 \\
& 0.4 & 2.30E1 & 9.5E-2 & 1.9E-12\\
& 0.6 &  4.96E2 & 1.44E1 &  3.6E-1\\
\tableline
1.0 & 0.0 & 1.43E2 & 7E-4 & 0.0E0 \\
& 0.1& 2.51E2 & 4.1E-2 & 6E-11\\
& 0.2& 3.87E2 & 6.52E0 & 0.0E0\\
& 0.4 & 7.33E2 & 2.24E2 & 1.05E1\\
& 0.6 & 1.95E3 & 6.82E2 & 3.04E2\\
\tableline
3.0 & 0.0 & 6.46E2 & 1.32E2 & 6.6E-2\\
& 0.1& 9.78E2 & 2.03E2 & 3.01E1\\
& 0.2& 1.45E3 & 3.04E2 & 1.04E2\\
& 0.4 &  2.99E3 & 9.07E2 & 3.22E2\\
& 0.6 & 7.12E3 & 2.60E3 & 1.19E3\\
\tableline
10 & 0.0 &  2.48E3 & 6.01E2 & 1.92E2\\
& 0.1&  3.28E3 & 9.36E2 & 3.01E2\\
& 0.2& 4.64E3 & 1.42E3 & 6.45E2 \\
& 0.4 &  9.27E3 & 3.05E3 & 1.45E3\\
& 0.6 & 2.31E4 & 7.84E3 & 3.82E3 \\
\tableline     
\end{tabular}
\end{center}
For $\chi = 45^0$, $\phi = 270^0$, $\xi = 0.5$ and neutron star parameters $M_\sun = 2.15$, $R_{\rm ns} = 9.9$ km and $I = 1.56 \times 10^{45}\,\rm g\,cm^2$.
\end{table}
\newpage
      

\newpage
\begin{figure}
\includegraphics[width=180mm]{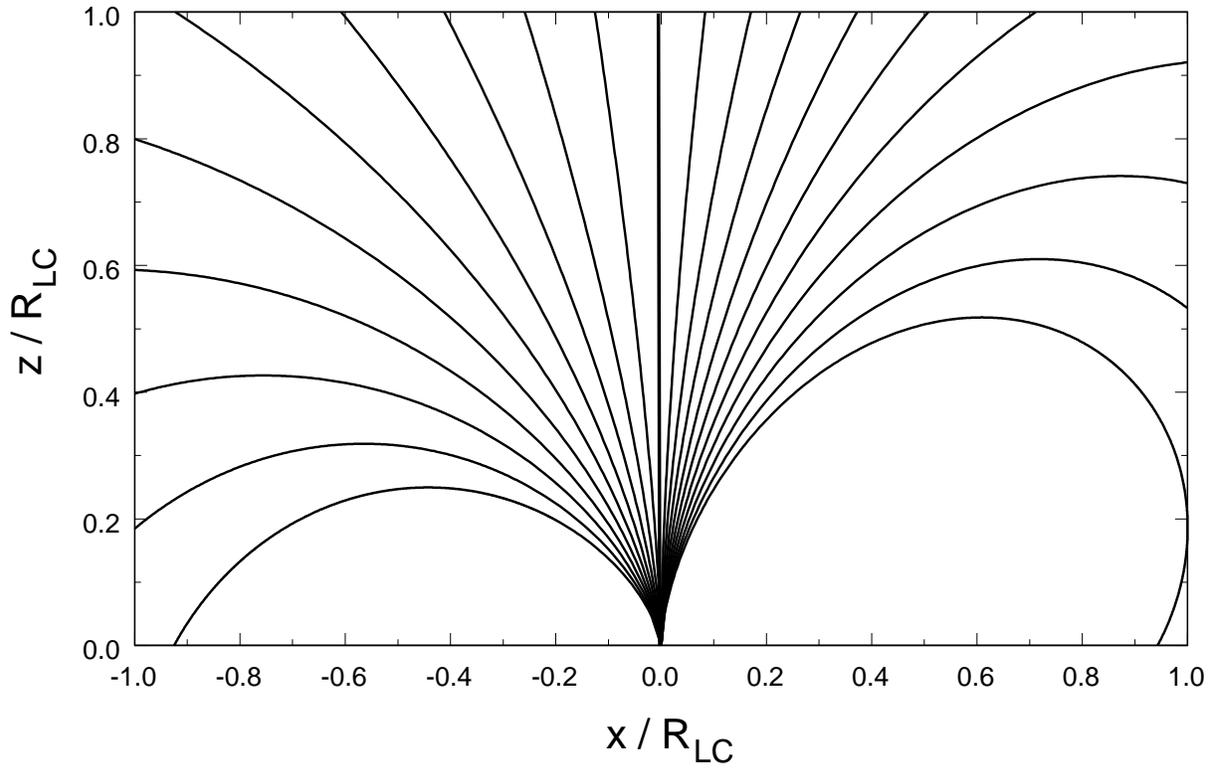}
\caption{Field lines of distorted magnetic dipole having an offset polar cap in the x-z plane and offset parameter $\epsilon = 0.2$.}    
\end{figure}

\newpage 
\begin{figure}
\hspace{-1.0cm}
\includegraphics[width=210mm]{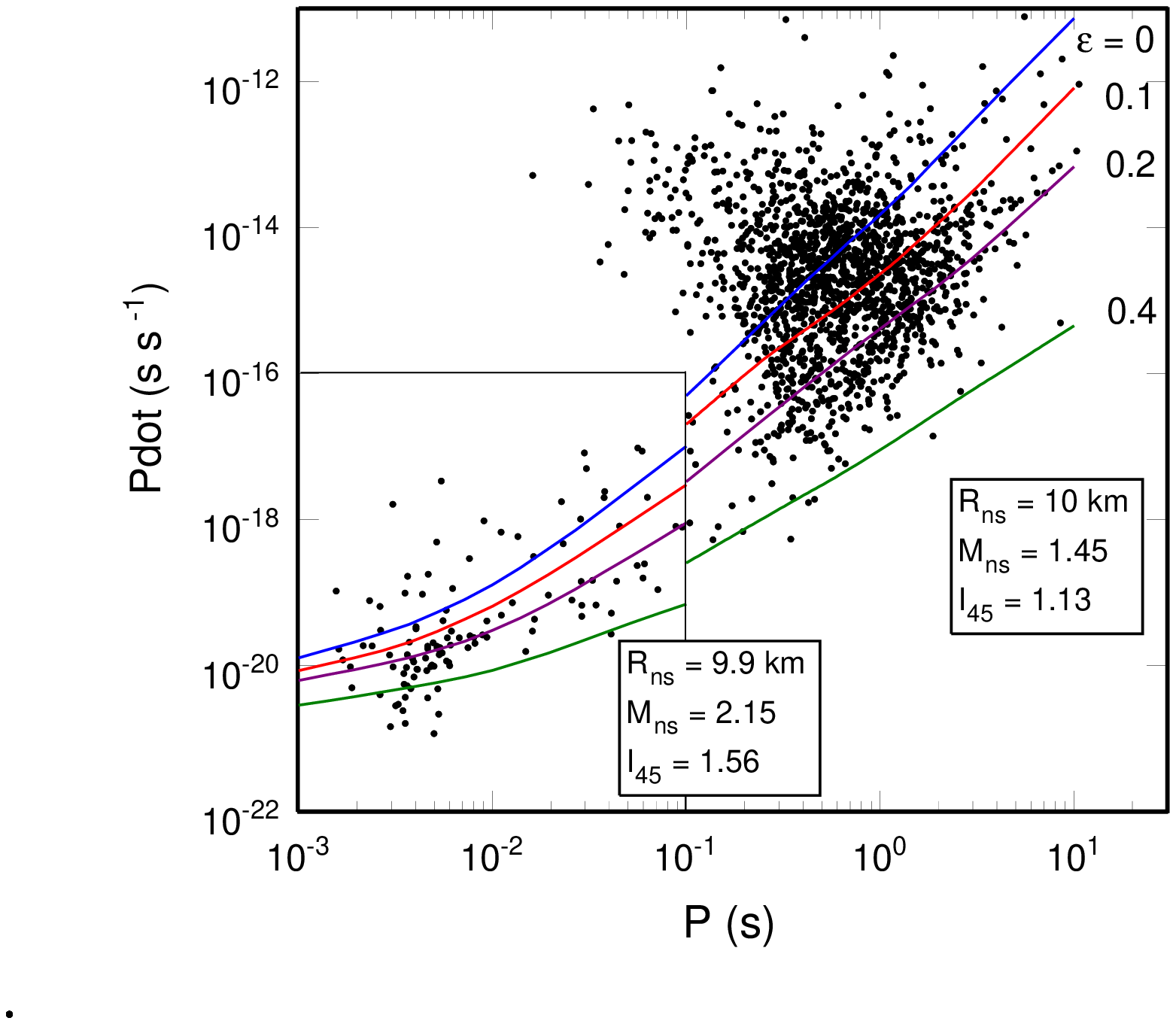}
\caption{Death lines  in the $P$-$\dot P$ diagram for pair production by curvature radiation, for different values of the offset parameter $\epsilon$ and inclination angle $\chi = 60^\circ$.  The NS radius $R_{\rm ns}$, mass $M_{\rm ns}$ (in Solar mass units) and moment of inertia $I_{45} = I/10^{45}\,\rm g\, cm^2$ refer to different NS  equations of state used for normal and millisecond pulsars and are described in the text.  Radio pulsars with measured $\dot P$ from the ATNF catalog (Manchester et al. 2005, http://www.atnf.csiro.au/research/pulsar/psrcat) are plotted as black dots. }    
\end{figure}


\clearpage
\begin{references}
\reference{}
Abdo, A. A. et al. 2009, Science, 325, 848.
\reference{} 
Abdo, A. A. et al. 2010, ApJ, 712, 957
\reference{}
Arons, J. 1983, ApJ, 266, 215.
\reference{}
Arons, J. 1997, in Neutron Stars and Pulsars: Thirty Years after the Discovery, eds. N. Shibazaki, N. Kawai, S. Shibata and T. Kifune, University Academy Press, Inc., Tokyo, Japan, p. 339.
\reference{} 
Arons, J.; Backer, D. C.; Spitkovsky, A.; Kaspi, V. M. 2005, in Binary Radio Pulsars, ASP Conference Series, Vol. 328,  Ed.  F. A. Rasio and I. H. Stairs (San Francisco: Astronomical Society of the Pacific) p.95
\reference{}
Arons, J. \& Scharlemann, E. T. 1979, ApJ, 231, 854
\reference{}
Bai, X-N. \& Spitkovsky, A. 2010, ApJ, 715, 1282.
\reference{}
Baring, M. G. \& Harding, A. K. 2001, ApJ, 547, 929.
\reference{}
Beloborodov, A. M. 2008, ApJ, 683, L41.
\reference{}
Bogdanov, S. , Rybicki, G. B. \& Grindlay, J. E. 2007, ApJ, 670, 668.
\reference{}
Bogdanov, S. \& Grindlay, J. E. 2009, ApJ, 703, 2259.
\reference{}
Cheng, K.~S., Ho, C., \& Ruderman, M.~A. 1986, ApJ, 300, 500.
\reference{}
Contopoulos, I., Kazanas, D. \& Fendt, C. 1999, ApJ, 511, 351.
\reference{}
Daugherty, J. K. \& A. K. Harding 1982, ApJ, 252, 337.
\reference{}
Daugherty, J. K. \& A. K. Harding 1996, ApJ, 458, 278.
\reference{} 
De Jager, O. C. 2007, ApJ, 658, 1177.
\reference{}
De Jager, O. C., A. K. Harding, Michelson, P. F. , Nolan, P. L., Sreekumar, P. \& Thompson, D. J. 1996, ApJ, 457, 253.
\reference{}
Deutsch, A. J., 1955, Ann. d'Astrophys., 18, 1.
\reference{}
Dyks, J. \& Harding, A. K. 2004, ApJ, 614, 869.
\reference{}
Friedman, J. L., Ipser, J. R., \& Parker, L. 1986, ApJ, 304, 115.
\reference{}
Harding, A.~K., \& Muslimov, A.~G. 1998, ApJ, 508, 328.
\reference{}
Harding, A.~K., \& Muslimov, A.~G. 2001, ApJ, 556, 987.
\reference{}
Harding, A.~K., \& Muslimov, A.~G. 2002, ApJ, 568, 862.
\reference{}
Hibschman, J. A. \& Arons, J. 200, ApJ, 560, 871.
\reference{}
Hirotani, K. 2008, Open Astronomy (arXiv:0809.1283)
\reference{}
Lamb, F. K.  et al, 2009, ApJ, 706, 417.
\reference{}
Lattimer, J. M. \& Prakash, M. 2007, Physics Reports, 442, 109.
\reference{}
Luo, Q.; Shibata, S.; Melrose, D. B. 2000, MNRAS, 318, 943.
\reference{}
Lyutikov, M. 2004, MNRAS, 353, 1095.
\reference{}
Manchester, R. N., Hobbs, G. B., Teoh, A. \& Hobbs, M. 2005, Astron. J., 129, 1993
\reference{}
Medin, Z. \& Lai, D. 2010, MNRAS, 406, 1379.
\reference{}
Muslimov, A. G. \& Tsygan, A. I. 1992, MNRAS, 255, 61 
\reference{}
Muslimov, A. G. \& Harding, A. K.  2004, ApJ, 606, 1143
\reference{}
Putney, A. \& Jordan, S. 1995, ApJ, 449, 863.
\reference{}
Ruderman, M. 1991, ApJ, 366, p. 261. 
\reference{}
Ruderman, M.A. \& Sutherland, P. G. 1975, ApJ, 196, 51 
\reference{}
Russell, C. T.  \& Daugherty, M. K. 2010,  Space Science Reviews, 152,  251.
\reference{}
Spitkovsky, A. ApJ, 648, L51 (2006)
\reference{}
Sturrock, P. A. 1971, ApJ, 164, 529.
\reference{}
Timokhin, A. MNRAS, 36, 1055 (2006)
\reference{}
Venter, C.; Harding, A. K.; Guillemot, L. 2009, ApJ, 707, 800.
\reference{}
Wickramasinghe, D. 2001, in Magnetic Fields Across the Hertzsprung-Russell Diagram, AIP Conference Series, Vol. 248, eds. G. Mathys, S. K. Solanki, and D. T. Wickramasinghe.
\reference{}
Young, M. D., Manchester, R. N. \& Johnston, S. 1999, Nature, 400, 848

\end{references}
\end{document}